\documentstyle[twoside,fleqn,espcrc2,epsf]{article}

\newcommand{\AmS}{{\protect\the\textfont2
  A\kern-.1667em\lower.5ex\hbox{M}\kern-.125emS}}

\title{
\vspace*{-35pt}
{\normalsize \hfill {\sf UTCCP-P-44}} \\
\vspace*{-6pt}
{\normalsize \hfill {\sf Aug.\ 1998}} \\
Non-perturbative determination of anisotropy coefficients
and pressure gap at the deconfining transition of QCD
\thanks{presented by S.\ Ejiri
}}

\author{S.\ Ejiri\rlap,\address{Center for Computational Physics, 
University of Tsukuba, Tsukuba, Ibaraki 305-8577 , Japan}
\addtocounter{address}{-1}
Y.\ Iwasaki\rlap,\addressmark
\addtocounter{address}{-1}
and K.\ Kanaya\addressmark
}

\begin{document}

\begin{abstract}
We propose a new non-perturbative method 
to compute derivatives of gauge coupling constants 
with respect to anisotropic lattice spacings (anisotropy coefficients).
Our method is based on a precise measurement of the finite temperature 
deconfining transition curve in the lattice coupling parameter space 
extended to anisotropic lattices 
by applying the spectral density method.
We determine the anisotropy coefficients 
for the cases of $SU(2)$ and $SU(3)$ gauge theories.
A longstanding problem, 
when one uses the perturbative anisotropy coefficients, 
is a non-vanishing pressure gap at the deconfining transition point 
in the $SU(3)$ gauge theory.
Using our non-perturbative anisotropy coefficients, we find that
this problem is completely resolved. 
\end{abstract}

\maketitle

\section{Introduction}
\label{sec:intro}

In a phenomenological study of heavy ion collisions and 
evolution of early Universe, 
it is important to evaluate 
the energy density and the pressure of the quark-gluon plasma
near the transition temperature of the deconfining phase transition of QCD. 

On an anisotropic lattice with $a_s$ and $a_t$ the lattice spacings 
in spatial and temporal directions, 
the standard plaquette action for $SU(N_c)$ gauge theory 
is given by 
$ 
 S = -\beta_s \sum P_{s}(x)-\beta_t \sum P_{t}(x),
$ 
where $ P_{s(t)}$ is the spatial (temporal) plaquette. 
Hence the energy density, 
$\epsilon=-\frac{1}{V} \frac{\partial \ln Z}{\partial T^{-1}}$, 
and the pressure, 
$p=T\frac{\partial \ln Z}{\partial V}$, 
are expressed in terms of the anisotropy coefficients, 
$a_t \frac{\partial \beta_s}{\partial a_t},
a_t \frac{\partial \beta_t}{\partial a_t},
\frac{\partial \beta_s}{\partial \xi},
\frac{\partial \beta_t}{\partial \xi}.$
We choose $a_t$ and $\xi \equiv a_s/a_t$ as 
independent variables to vary the lattice spacings. 

Perturbative values for these anisotropy coefficients were calculated
by Karsch in \cite{karsch}.
However, when we apply them to data obtained by MC simulations,
we encounter pathological results such as a negative pressure 
and a non-vanishing pressure gap at the deconfining transition point
of $SU(3)$ gauge theory.
Non-perturbative anisotropy coefficients are, therefore, 
required to study $\epsilon$ and $p$ in MC simulations. 

Two non-perturbative methods have been adopted 
to determine the anisotropy coefficients.
One is ``the matching method'' \cite{burgers,scheideler,klassen} 
based on a measurement of $\xi$ as a function of $\beta_s$ 
and $\beta_t$ by matching spatial and temporal Wilson loops. 
The other is a method based on a non-perturbative estimate of 
pressure obtained by ``the integral method'' \cite{engels,boyd}.

In this paper, 
we propose a new non-perturbative approach to compute 
the anisotropy coefficients,
and determine the coefficients 
for $SU(2)$ and $SU(3)$ gauge theories \cite{ejiri}.
We restrict ourselves to the case of 
isotropic lattices, $\beta_s\;=\;\beta_t\;\equiv\;\beta$,
where most simulations are performed.
In this case, two anisotropy coefficients 
are just the beta-function at $\xi \;=\; 1$;
$(a_t \frac{\partial \beta_s}{\partial a_t})_{\xi = 1}
= (a_t \frac{\partial \beta_t}{\partial a_t})_{\xi = 1}
\equiv a \frac{{\rm d} \beta}{{\rm d} a} $, 
whose non-perturbative values are well studied both in $SU(2)$ and
$SU(3)$ gauge theories 
\cite{taro,engels,boyd,edwards}.
Furthermore, a combination of the remaining two anisotropy coefficients 
is known to be again related to the beta-function by 
$
\left(\frac{\partial \beta_s}{\partial \xi} 
+ \frac{\partial \beta_t}{\partial \xi}\right)
_{a_t : {\rm fixed}} 
= \frac{3}{2} \, a \frac{{\rm d} \beta}{{\rm d} a} 
$
\cite{karsch}. 
Therefore, only one additional input is required to determine 
the anisotropy coefficients for isotropic lattices.

\section{Method}
\label{sec:method}

Our method is based on an observation that, 
the transition temperature $T_{c} = 1/\{N_t a_t (\beta_s, \beta_t)\}$ 
is independent of the anisotropy of the lattice. 
This brings us the following relation between 
the anisotropy coefficients and 
the slope $r_t$ of the transition curve in the $(\beta_s, \beta_t)$ plane
at $\xi = 1$, 
\begin{eqnarray}
r_t 
\equiv \left. \frac{{\rm d} \beta_s}{{\rm d} \beta_t} 
\right|_{\;\rm trans.curve}
= \left(\frac{\partial \beta_s}{\partial \xi}\right)_{\;\xi = 1} 
\left/ \left(\frac{\partial \beta_t}{\partial \xi}\right)_{\;\xi = 1} \right. .
\label{eqn:rt}
\end{eqnarray}
From this equation,
we obtain the expressions for the customarily used forms 
for the anisotropy coefficients 
$c_{s (t)} \;=\; ( {\partial g_{s (t)}^{-2} }/
    {\partial \xi} )_{a_s: {\rm fixed}}$ 
where $\beta_{s}=2N_{c} g_{s}^{-2} \xi^{-1}$ and 
$\beta_{t}=2N_{c} g_{t}^{-2} \xi$;
\begin{eqnarray}
c_s 
&=& \frac{1}{2N_{c}} \left\{ \beta + \frac{r_t - 2}{2 ( 1 + r_t)} \,
    a \frac{{\rm d} \beta}{{\rm d} a} \right\}, \nonumber \\ 
c_t 
&=& \frac{1}{2N_{c}} \left\{ - \beta 
    + \frac{1 - 2 r_t}{2 ( 1 + r_t)} \, 
    a \frac{{\rm d} \beta}{{\rm d} a} \right\}.
\label{kc}
\end{eqnarray}
Therefore, when the value for the beta-function is available, 
we can determine these anisotropy coefficients 
by measuring $r_t$ from the finite temperature transition 
curve in the $(\beta_s, \beta_t)$ plane. 

As a result, $\epsilon$ and $p$ are given by 
\begin{eqnarray}
\frac{\epsilon-3p}{T^4} \hspace{-2mm} &=& \hspace{-2mm}
- 3 N_t^4 \, a \frac{{\rm d}\beta}{{\rm d}a} \,
\{\langle P_s \rangle + \langle P_t \rangle\ - 2\langle P \rangle_0\}, 
\label{eq:e3p} \\
\frac{\epsilon+p}{T^4} \hspace{-2mm} &=& \hspace{-2mm} 
3 N_t^4 \, a \frac{{\rm d}\beta}{{\rm d}a} \, 
\frac{r_t-1}{r_t+1} \,
\{\langle P_s \rangle - \langle P_t \rangle\},
\label{eq:emp}
\end{eqnarray}
where $\langle P \rangle_{0} $ is the plaquette at $T=0$. 

In order to determine the transition curve in the $(\beta_s, \beta_t)$ plane,
we compute the rotated Polyakov loop $L$.
We define the transition point as the peak position of 
the susceptibility 
$\chi = N_s^3(\langle L^2 \rangle - \langle L \rangle^2).$ 
The coupling parameter dependence of $\chi$ in the 
$(\beta_s,\beta_t)$ plane is computed 
by applying the spectral density method \cite{swendsen} 
extended to anisotropic lattices. 
This enables us to compute the anisotropy coefficients 
directly from simulations at $\xi \approx 1$ 
without introducing an interpolation Ansatz, 
unlike the case of previous studies. 

\section{Results}

\begin{figure}[tb]
\vspace*{-5mm}
\centerline{
\epsfxsize=6cm\epsfbox{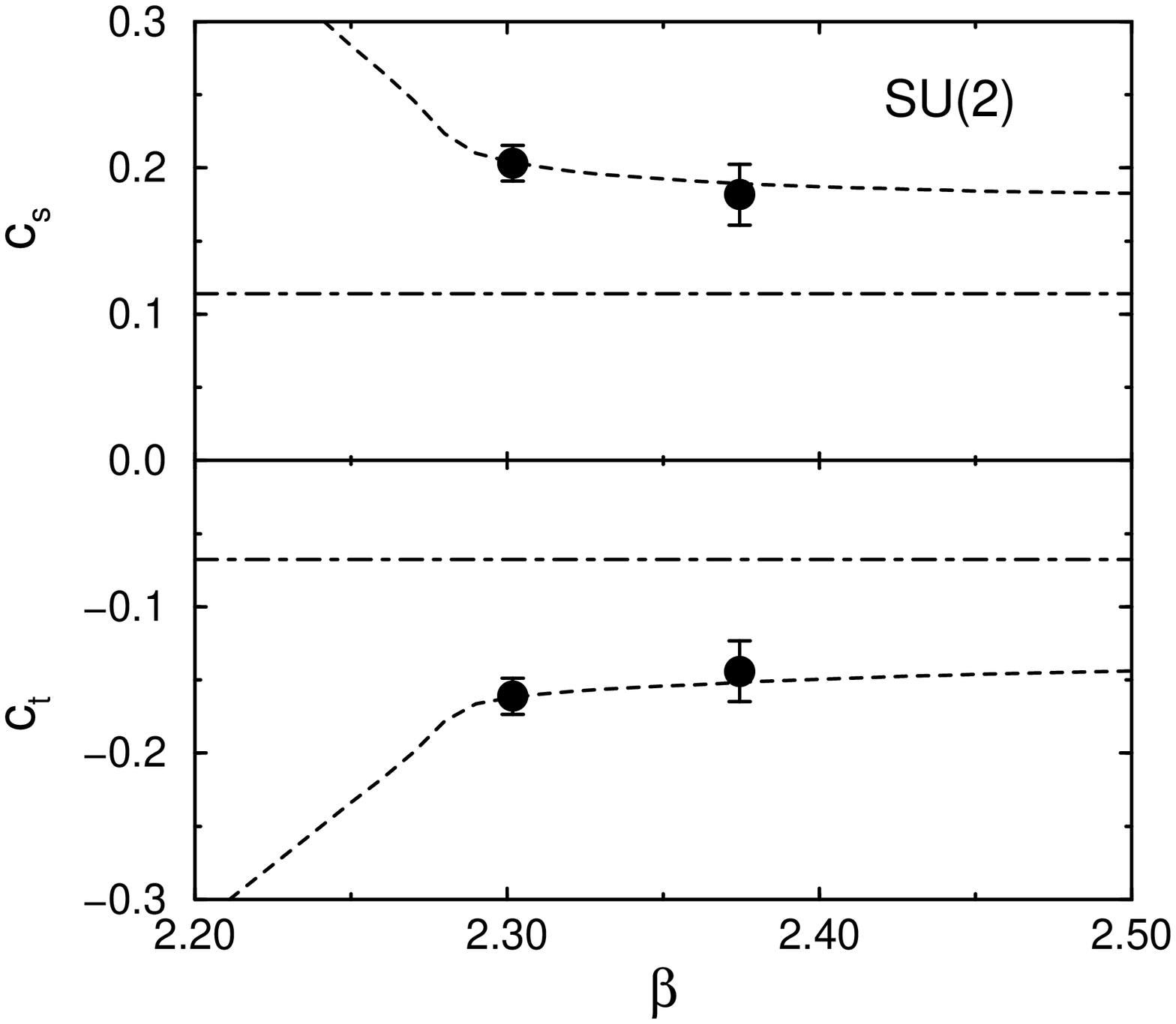}
}
\vspace*{-24mm}
\centerline{
\epsfxsize=6cm\epsfbox{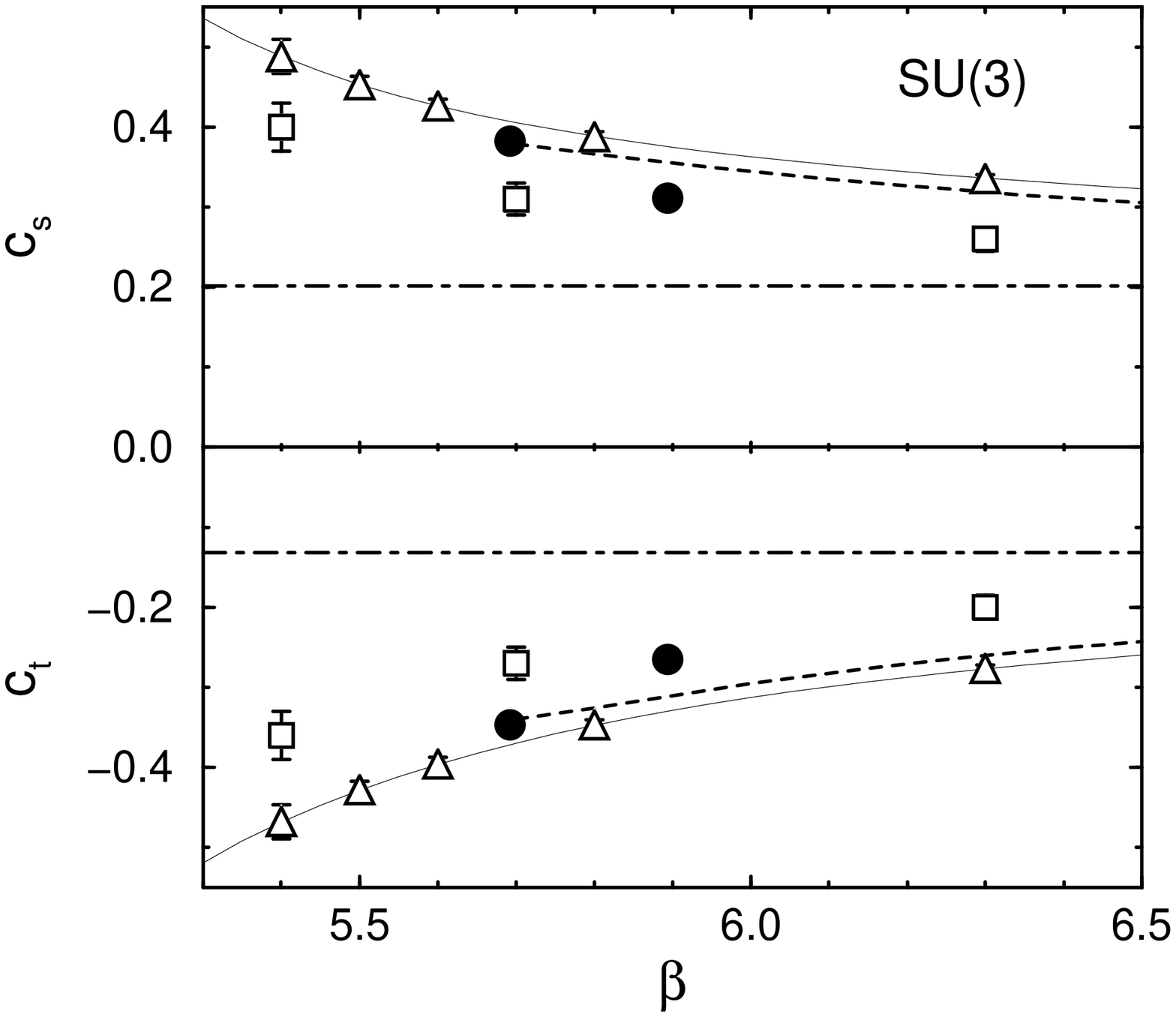}
}
\vspace*{-30mm}
\caption{
Anisotropy coefficients $c_s$ and $c_t$ 
for the $SU(2)$ and $SU(3)$ gauge theories.
}
\label{fig:kc}
\end{figure}

\begin{figure}[tb]
\vspace*{-12mm}
\hspace*{-8mm}
\centerline{
\epsfxsize=7.2cm\epsfbox{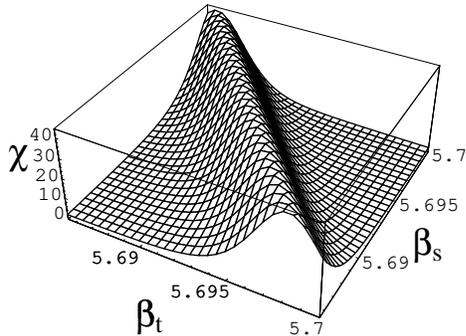}
}
\vspace*{-12mm}
\caption{
The Polyakov loop susceptibility for $SU(3)$ 
obtained on a $24^3\times36\times4$ lattice.
}
\label{fig:su3sus}
\end{figure}

\begin{table}[tb]
\caption{Latent heat, pressure gap, 
$ \Delta \langle P_t \rangle \left/ \Delta \langle P_s \rangle \right.$
and $-r_{t}$ at the deconfining transition point of $SU(3)$ gauge 
theory.
}
\label{tab1}
\begin{center}
\begin{tabular}{lcc}
\hline
lattice            & $24^2\times36\times4$ & $36^2\times48\times6$ \\
$\beta$                  &   5.6925     &    5.8936     \\
\hline
$\Delta \epsilon/T^4$    &   2.074(34)  &   1.569(40) \\
$\Delta p/T^4$           &   0.001(15)  & $-$0.003(17) \\
$\Delta \langle P_t \rangle \left/ \Delta \langle P_s \rangle \right.$
                         &   1.201(35)  &   1.218(46) \\
$- r_t$                  &   1.201(1)   &   1.220(3)  \\
\hline
\end{tabular}
\end{center}
\end{table}

We first test the method for the case of $SU(2)$ gauge theory 
at the transition point $\beta_c$ for $N_t=4$ and 5. 
Simulations are performed on $ 16^{3} \times 4 $ and $20^3\times5$ lattices.
Results for $c_s$ and $c_t$ are denoted by filled circles 
in Fig.~\ref{fig:kc} (top).
Our results are consistent with the results from the integral method
(doted curves) \cite{engels}.

We then study the more realistic case of the $SU(3)$ gauge theory.
Because the method works well even with data obtained only 
on isotropic lattices, 
we analyze the high statistics data by the QCDPAX Collaboration \cite{QCDPAX}.
Simulations were performed at the deconfining transition point for
$N_t=4$ and 6 
on five lattices.
Details of the $SU(3)$ simulations are given in \cite{QCDPAX}.

Fig.~\ref{fig:su3sus} shows the $(\beta_s,\beta_t)$ dependence of
the susceptibility on a $24^2\times 36\times 4$ lattice.
Because the peak of the susceptibility becomes sharper
as the spatial volume of the lattice is increased, 
we can measure $r_t$ most precisely on the spatially largest 
lattices. 
Therefore, in the following, 
we use the results obtained on the largest 
$24^2\times36\times4$ and $36^2\times48\times6$ lattices.
The values obtained on smaller lattices are consistent.
For the beta-function, we adopt a result computed from a 
recent string tension data \cite{edwards}.

In Fig.~\ref{fig:kc} (bottom), we summarize our results for the 
$c_{s}$ and $c_{t}$ of the $SU(3)$ gauge theory (filled circles) 
together with previous values: 
the perturbative results (dot-dashed lines) \cite{karsch},
results from the integral method (doted curves) \cite{boyd},
and those from the matching of Wilson loops on anisotropic lattices
(squares \cite{scheideler}, triangles \cite{klassen}).
We find that all non-perturbative methods give values which 
are roughly consistent with each other, showing a clear
deviation from the perturbation theory. 

The deconfining transition is of first order for $SU(3)$.
At a first order transition point, we have a finite gap for energy 
density, the latent heat, but expect no gap for pressure.
It is known that the perturbative anisotropy coefficients leads to 
a non-vanishing pressure gap at the deconfining transition
point: $\Delta p / T^4 = -0.32(3)$ and $-0.14(2)$ at
$N_t=4$ and 6 \cite{QCDPAX}.

New values for the gaps in $\epsilon$ and $p$ using our non-perturbative 
anisotropy coefficients are summarized in Table \ref{tab1}.
We find that the problem of non-zero pressure gap is completely
resolved with our non-perturbative anisotropy coefficients.

We note that, 
because the beta-function appears only as a common overall factor
in (\ref{eq:e3p}) and (\ref{eq:emp}), 
the conclusion that $\Delta p$ vanishes with our anisotropy coefficients
does not depend on the value of the beta-function.
Actually we have from eqs.(\ref{eq:e3p}) and (\ref{eq:emp})
a simple condition for $\Delta p = 0$: 
\begin{eqnarray}
\Delta \langle P_t \rangle \left/ \Delta \langle P_s \rangle \right.  
= - r_{t}
\label{eq:cnddp0}
\end{eqnarray}
where $\Delta \langle P_{s(t)} \rangle$ is the gap 
in the spatial (temporal) plaquette between the two phases. 
Although the two sides of (\ref{eq:cnddp0}) 
are obtained from quite different measurements, 
they agree precisely with each other as shown in Table~\ref{tab1}.

\vspace{2mm}

We are grateful to J.\ Engels, T.R.\ Klassen and 
the members of CCP, Tsukuba for useful discussions.
This work is in part supported by 
the Grants-in-Aid of Ministry of Education,
Science and Culture (Nos.~08NP0101, 09304029, 10640248). 
SE is supported by JSPS.

\end{document}